\documentclass[12pt,letterpaper]{article}

\usepackage[margin=1in]{geometry}
\usepackage[T1]{fontenc}
\usepackage[utf8]{inputenc}
\usepackage{lmodern}
\usepackage{amsmath,amssymb}
\usepackage{newunicodechar}
\newunicodechar{Ω}{$\Omega$}
\newunicodechar{∆}{$\Delta$}
\newunicodechar{π}{$\pi$}
\newunicodechar{⋆}{$\star$}
\usepackage{microtype}
\usepackage{booktabs,tabularx,array}
\usepackage{enumitem}
\usepackage[numbers,sort&compress]{natbib}
\usepackage[hidelinks]{hyperref}

\newcolumntype{Y}{>{\raggedright\arraybackslash}X}
\newcommand{\oibar}{\textsc{O-I-B-A-R}}
\setlist{nosep,leftmargin=2.2em}
\title{Why Organizational Rules Fail AI: O-I-B-A-R and the\\Externalization of Decision Boundaries}
\author{\begin{tabular}{cc}
Chao Li & Chunyi Zhao\\
\texttt{lichao@aegt.org} & \texttt{chunyizhao@aegt.org}
\end{tabular}}
\date{August 2026}

\begin{document}
\nocite{r1,r2,r3,r4,r5,r6,r7,r8,r9,r10,r11,r12,r13,r14,r15,r16,r17,r18,r19,r20,r21,r22,r23,r24,r25,r26}
\maketitle
\begin{abstract}
AI systems increasingly enter organizations through policies, standard operating procedures, expert playbooks, prompts, and other explicit representations of work. Organizational research has long shown that formal descriptions differ from situated practice. Recent studies of AI in knowledge work likewise show that formally captured ``know-what'' can omit the contextual know-how experts use when judgments become uncertain. We argue that a recurring class of organizational AI failures arises partly from a knowledge-representation problem at the sociotechnical interface, alongside limitations in model capability: the AI receives the procedure, while the organization operates on the procedure plus its negative boundaries, runtime judgments, responsibility assignments, and learning history. We introduce \oibar{} (OPEN, IS, BUT, ACTION, RESULT), a cognitive scaffold for externalizing these missing decision boundaries. IS records when a judgment holds; BUT records a concrete failure that contains information beyond the logical negation of IS. Comparable success/failure cases are decomposed toward a minimally sufficient changing variable, which becomes a value-bearing decision dimension. A suspension represents the intermediate state in which the dimension is known but its value for the present case is unresolved; this state specifies what must be measured, asked, retrieved, or escalated to a human. RESULT then confirms a boundary, shifts a threshold, or exposes a new dimension. We position the method in relation to Kelly's repertory-grid elicitation, Winston's near-miss learning, Mitchell's version spaces, cognitive task analysis, situated-action and organizational-routine research, and counterexample-guided refinement. Contrastive elicitation and tacit-knowledge externalization have clear precedents. Our narrower contribution is their integration into an open-world organizational decision representation in which incidents can generate new dimensions, unresolved values define human--AI handoffs, and feedback can expand the represented decision space. We also identify a sociotechnical tension: durable and attributable failure histories can suppress the candor on which useful boundary knowledge depends. Externalization must therefore be designed as an organizational intervention with real costs and incentives.
\end{abstract}

\section{Introduction}

AI deployment in organizations creates a puzzle. Systems that perform impressively on well-specified tasks can still fail in ways that experienced workers describe as obvious: following a rule when an exception was salient, acting when a question should have been asked, escalating the wrong cases, or applying a formal criterion while missing the contextual cue on which practitioners actually rely. Research on algorithms and organizations has repeatedly emphasized that AI reshapes expertise, occupational boundaries, coordination, control, and the division of labor; its organizational effects extend well beyond the automation of isolated tasks [1--3]. Field studies of AI in professional work likewise show that formally encoded labels and outputs may fail to capture the know-how experts use in practice. In one hospital study, multiple AI tools with strong benchmark-style accuracy failed to meet organizational expectations until managers confronted the gap between experts' know-what and their richer know-how \cite{r4}.

This problem has a deeper organizational lineage. Suchman argued that plans serve as resources for situated action and do not determine action in detail \cite{r5}. Brown and Duguid showed that the ways people actually work can differ fundamentally from manuals, training programs, organizational charts, and job descriptions, and that ``noncanonical'' practice often carries the learning and innovation on which work depends \cite{r6}. Orr's ethnography of copier technicians showed expert diagnosis being sustained through situated improvisation and context-rich ``war stories'' that preserved cases official documentation did not capture \cite{r7}. Feldman and Pentland later distinguished the ostensive description of an organizational routine from its concrete performative enactments, warning against treating the abstract summary of work as the work itself \cite{r8}.

Attempts to make these missing boundaries machine-readable expose a second sociotechnical problem. Failure knowledge may be strategically withheld even when it is absent from formal systems. Research on organizational silence describes conditions under which employees collectively come to view speaking about problems as unwise \cite{r24}, while work on psychological safety links interpersonal risk to whether teams engage in learning behavior \cite{r25}. Studies of organizational failure likewise show that front-line workarounds can restore local performance while preventing deeper learning from recurring problems \cite{r26}. Any externalization method must therefore preserve the social conditions under which truthful failure knowledge is produced.

These literatures suggest a specific diagnosis for organizational AI. The problem is often framed as if an AI system were given the organization's knowledge and then failed to use it with sufficient intelligence. In many deployments, however, the organization has given the system only the canonical positive rule: what normally should be done. Experienced practice additionally depends on where that rule breaks, which variable determines the crossing of that boundary, what information must be acquired at runtime, and who is responsible for acquiring it. The diagnosis can be summarized as follows:

\begin{quote}
\emph{The AI receives the procedure; the organization operates on the procedure plus its exceptions, judgments, and repair history.}
\end{quote}

We call the missing structure decision-boundary knowledge. The central research question of this paper is:

\begin{quote}
\emph{How can organizations externalize the boundary conditions, runtime judgments, and responsibility structures that govern real-world decisions, so that AI systems need not treat explicit procedures as universally applicable rules?}
\end{quote}

We introduce \oibar{}, a five-stage scaffold for this purpose. The method begins with a judgment in a concrete domain (OPEN), records a positive applicability condition (IS) and a concrete failure condition (BUT), compares the two to identify a decision-relevant changing variable, turns the unresolved value of that variable into an actionable query (a suspension), commits the current model to action (ACTION), and uses the resulting evidence (RESULT) to confirm, refine, or expand the represented decision space.

Prior work constrains the paper's novelty claim. Contrastive concept learning is old: Kelly's repertory-grid method elicits constructs through similarity and difference without imposing a fixed questionnaire taxonomy \cite{r9}; Winston's near-miss learning uses carefully chosen positive and negative examples to reveal critical structural distinctions \cite{r10,r11}; Mitchell's version spaces refine concept boundaries using positive and negative examples within a specified hypothesis language \cite{r12}; and cognitive task analysis has long elicited expert cues and decision requirements from critical incidents \cite{r19,r20}. \oibar{} does not claim to invent any of these ideas. Its proposed contribution integrates four elements that become consequential in open-world organizational AI: (i) failure incidents recalled from practice, as distinct from teacher-curated training sets; (ii) discovery and domain-relative naming of value-bearing decision dimensions without a fixed feature vocabulary; (iii) a typed unknown state in which a dimension is known but its current value is unresolved; and (iv) a feedback loop in which results can add new dimensions, while updates in a closed representation remain limited to existing values.

\textbf{Contributions.} The paper makes six concrete contributions:

\begin{enumerate}[leftmargin=2.6em]
\item A sociotechnical diagnosis of rule failure. We separate four representational gaps in organizational AI deployment: the boundary gap, runtime-value gap, responsibility gap, and learning gap.
\item Asymmetric boundary representation. IS and BUT are not logical complements: IS captures an applicability region, while BUT is a concrete failure capable of introducing decision-relevant information absent from the original rule.
\item Decision-dimension discovery. A qualified IS/BUT pair is decomposed toward a minimally sufficient changing variable. This sharpens the familiar one-difference intuition of near-miss learning into an elicitation target without presenting it as a new learning principle.
\item Suspension as a human--AI handoff primitive. Once a dimension is known but its value in the current case is not, the framework represents the missing value explicitly, identifying what must be queried and, when it cannot be resolved automatically, what a human is substantively being asked to judge.
\item Domain-relative dimension naming. The framework treats question patterns as primary and uses the domain to constrain where abstraction should converge; noun hypernyms do not serve this function.
\item Feedback-driven growth and bounded stability monitoring. RESULT distinguishes threshold revision from structural expansion: a failure that cannot be expressed by the current dimensions can introduce a new dimension. Apparent stability is treated as evidence about the current exposure process, not as proof that the latent dimension set is complete.
\end{enumerate}

The aim is not to propose a universal theory of cognition. It is to offer a representational mechanism that can be inspected, criticized, implemented, and empirically tested in organizational human--AI systems.

\section{Why Organizational Rules Fail AI: Four Representational Gaps}

The gap between canonical descriptions and situated practice becomes particularly consequential when organizational text is treated as executable guidance. We distinguish four gaps that are often collapsed into the vague claim that an AI system lacks ``judgment.''

\subsection{Boundary gap: the organization writes the positive rule}

A standard operating procedure typically specifies what to do under an intended condition. It rarely preserves the equally important incident in which the same action failed. Brown and Duguid's distinction between canonical descriptions and noncanonical work practice \cite{r6}, Orr's observations of story-based troubleshooting \cite{r7}, and the ostensive/performative distinction in organizational routines \cite{r8} all imply that the formal representation is only one side of work.

For AI, this creates a direct failure mode. If a policy says ``when a customer is upset, empathize before discussing a solution,'' but does not encode the case in which repeated empathy aggravated a customer who wanted a concrete refund amount, the model must infer the missing boundary. When it infers incorrectly, the organization may call the error a lack of common sense even though the relevant exception was never represented.

We call this the boundary gap: the positive prescription is explicit, while the negative applicability boundary remains tacit, anecdotal, or socially distributed.

\subsection{Runtime-value gap: a relevant dimension may be known while its value is not}

Even a complete list of relevant dimensions does not settle a live case. A team may know that need type, ownership, or risk severity matters, yet still not know its value for the current instance. This is distinct from generic model uncertainty. It is a typed missing value: the decision axis is known, but the present coordinate is not.

We call this the runtime-value gap. It matters because it converts an undifferentiated instruction such as ``use judgment'' into an operational question: ``which value must be obtained before acting?'' The answer may be available through a database, a measurement, a clarifying question, or live interpretation.

\subsection{Responsibility gap: ``human in the loop'' does not specify what the human is for}

Human--AI research often advocates augmentation and rejects full replacement \cite{r2,r3}, but ``keep a human in the loop'' is not itself a task specification. If a critical variable can only be obtained through situated interpretation, the unresolved variable identifies the content of the human role.

For example, ``complex cases go to a human'' leaves complexity undefined. ``Escalate when the customer's need type cannot be reliably determined'' states what the human must resolve.

We call the missing mapping from unresolved information to accountable judgment the responsibility gap. It shifts the unit of analysis from occupations to decision variables: the analysis asks which jobs remain human and, more precisely, which values in which decisions remain non-reducible at deployment time.

\subsection{Learning gap: action-level patches fail to update the model}

Organizations learn from incidents, but the learning is often stored as another prohibition: ``do not do that again,'' ``add a guardrail,'' or ``escalate this case.'' Argyris and Schön's distinction between single- and double-loop learning already highlights the difference between correcting actions and revising governing assumptions \cite{r22}. Counterexample-guided abstraction refinement in formal verification provides a computational analogue: a counterexample can refine the abstraction after a failed trace, extending the response beyond rejection of that trace alone \cite{r23}.

\begin{table}[htbp]
\centering
\small
\begin{tabularx}{\textwidth}{@{}lYYY@{}}
\toprule
\textbf{Gap} & \textbf{What is missing} & \textbf{Typical symptom} & \textbf{OIBAR primitive} \\
\midrule
Boundary & where the explicit rule fails & mechanical rule following & BUT \\
Runtime value & current value of a known dimension & ``it depends'' / guessing & suspension \\
Responsibility & who must obtain or judge the missing value & vague human handoff & suspension + role \\
Learning & whether failure moves a boundary or expands the variable set & guardrail accumulation & RESULT \\
\bottomrule
\end{tabularx}
\caption{Four representational gaps in organizational AI deployment.}
\label{tab:gaps}
\end{table}
For organizational AI, the analogous question is whether a failure can be explained using the current decision variables. If it can, the boundary or threshold may need refinement. If it cannot, the representation may be missing a dimension. We call failure to make this distinction the learning gap.

This framing makes the role of \oibar{} narrower and more sociotechnically specific than ``making tacit knowledge explicit.'' Externalization is the application; boundary and dimension construction are the mechanism.

\section{The O-I-B-A-R Framework}

\subsection{OPEN: a judgment under a domain}

OPEN specifies (i) a judgment that could fail and (ii) the domain in which that judgment is to be tested. A topic label alone is insufficient. Let J denote a decision-relevant judgment and Ω a domain of application. We write

\[
O=(J,\Omega).
\]

A weak OPEN is ``customer service'' or ``small-class education.'' Such phrases name topics, not judgments. A stronger OPEN is ``for renewal customers disputing a price increase, emotional acknowledgment should precede solution discussion.'' The latter can be tested in a recognizable setting. The domain Ω constrains which distinctions are useful and where abstraction should stop; it serves a substantive role beyond contextual metadata. The same question pattern may stabilize under different names in different domains. ``How large is X relative to Y ?'' may correspond to scale in algorithm selection, share in economics, or dose intensity in a clinical context. Naming is therefore domain-relative, with no global taxonomy assumed.

\subsection{IS: the positive boundary}

IS states a condition under which the current judgment is expected to hold. Let S be a set of observable or interpretable conditions: $\mathrm{IS}: S\Rightarrow J$. The notation has an operational interpretation and carries no claim of logical necessity. IS records a current applicability boundary: ``when these conditions obtain, acting as if J is true is presently justified.''

\subsection{BUT: a generative negative boundary}

BUT records a concrete case or condition in which applying J produces a materially different or undesirable outcome. Crucially, $\mathrm{BUT}\ne\neg\mathrm{IS}$. The logical negation of IS is derived from IS itself and therefore adds no new explanatory variable by construction. If IS is ``$k/n$ is small,'' then ``$k/n$ is not small'' remains entirely within the same variable set. A useful BUT is an observed or testable failure whose description may contain an explanatory variable absent from the IS statement. Let V (X) denote the set of decision-relevant variables explicitly or implicitly invoked by statement X. A simple generativity heuristic is $V(\mathrm{BUT})\setminus V(\mathrm{IS})\ne\varnothing$. This condition is not sufficient for a good BUT: irrelevant new words can always be added. It is a screening heuristic. The stronger requirement is causal or decision relevance: the new variable must help explain why the same judgment crosses from an acceptable to an unacceptable region. This distinction matters because ordinary ``limitations'' often behave like ¬IS: ``the result may not generalize outside the sample'' says little about what actually changes outside the sample. A generative BUT says what happens, under what condition, and which previously omitted factor is implicated.

\subsection{Boundary pairing and the discovery of a decision dimension}

Suppose a positive and negative boundary are comparable instances of the same judgment. Let ∆(IS, BUT | Ω) denote the minimal set of decision-relevant variables whose changed values are required to explain the transition from the IS state to the BUT state in domain Ω. A qualified boundary pair aims for $\lvert\Delta(\mathrm{IS},\mathrm{BUT}\mid\Omega)\rvert=1$. If |∆| = 0, the BUT is usually only a restatement or negation of IS. If |∆| > 1, multiple dimensions have been mixed and the pair should be decomposed. When |∆| = 1, the unique element d is treated as the decision dimension exposed by the pair: $D(\mathrm{IS},\mathrm{BUT}\mid\Omega)=d$. This operational criterion does not guarantee that natural language will admit a unique decomposition. In practice, multiple explanatory candidates may exist, and domain expertise is required to decide which variable is minimally sufficient. The method searches for the variable after observing both boundaries; it does not require a fixed taxonomy in advance.

\subsection{Failure attribution: from the action to the judgment}

A useful intermediate step is to rewrite the failure in terms of a judgment error and avoid stopping at the action error. For example:

\begin{center}
\small
\begin{tabularx}{0.9\textwidth}{@{}YY@{}}
\toprule
\textbf{Action-level attribution} & \textbf{Dimension-level attribution} \\
\midrule
``Empathy was wrong'' & ``The customer's need type was misclassified'' \\
``The heap was wrong'' & ``The scale regime was misjudged'' \\
``Caching was wrong'' & ``The data-change frequency was underestimated'' \\
\bottomrule
\end{tabularx}
\end{center}
Action-level attribution supports avoidance: ``do not do that again.'' Dimension-level attribution supports future discrimination: ``check this variable before deciding.'' This step converts incident memory into transferable knowledge.

\subsection{Suspension: an actionable unknown}

Once a dimension d is known, the current case may still lack its value. We call this unresolved but well-typed uncertainty a suspension. For an instance x,

\[
q_d(x)=?
\]

where qd is a query whose answer determines or constrains the value of d for x. Examples include:

\begin{itemize}
\item ``What is $k/n$ for this workload?'' (scale)
\item ``Is the input batch or stream?'' (input form)
\item ``Who owns this object, and may it be mutated?'' (ownership)
\item ``Does this customer need acknowledgment or a concrete remedy?'' (need type)
\end{itemize}

A suspension differs from generic uncertainty such as ``it depends.'' It is actionable because the missing information is typed by a known dimension. A well-formed suspension should, in principle, be answerable by measurement, retrieval, questioning, observation, or explicit human judgment. This distinction also identifies human responsibilities. If the suspension can be resolved from a database, the system can retrieve it. If its value exists only in the live context and requires human interpretation, then the unresolved dimension specifies the substantive content of the human role. ``Escalate complex cases'' becomes ``escalate when need type cannot be reliably determined.''

\subsection{Domain-relative recursive naming: the domain as a convergence operator}

The name of a dimension should not be generated by climbing a noun taxonomy. From the noun k, an ontology may produce ``variable,'' ``symbol,'' or ``number,'' but not ``scale.'' The relevant semantic object is the question pattern induced by the suspension. Let q0 be the suspension question with instance-specific terms abstracted away, and let

\[
A_{\Omega}(q)
\]

denote one step of domain-relative abstraction: ``within domain Ω, what is this question asking about?'' We then form $q_{t+1}=A_{\Omega}(q_t)$. A dimension label is operationally stable when a value-bearing expression d⋆ reaches a local fixed point: $A_{\Omega}(d^{\star})=d^{\star}$. For example,

\[
\text{``$X$ relative to $Y$: how large?''}\to\text{``relative magnitude''}\to\text{``scale''}\to\text{``scale.''}
\]

Some dimensions converge in zero or one step. ``Is there a copy?'' may already stabilize as copying.

This fixed-point language should not be read as a claim that natural-language abstraction is globally deterministic. The operator depends on the chosen domain, the professional vocabulary available in that domain, and human or model judgment. The practical claim is narrower: the domain acts as a convergence constraint. Without it, abstraction tends to climb toward content-free containers such as ``property'' or ``factor''; with it, the process can stop at a term that remains abstract enough to transfer yet concrete enough to take values.

Domain dependence is expected and should not be mistaken for lexical instability. The same underlying variable can be named differently across professional vocabularies. Identity should be judged by the variable being measured, not by word equality. A useful type test is whether the candidate dimension naturally takes values. Can one ask ``how large?'', ``which type?'', ``how many?'', or ``whose?'' A term such as ``important factor'' or ``relevant attribute'' often fails this test; it names a conceptual container, while a decision axis takes values.

\subsection{ACTION: committing the current model to reality}

ACTION is the decision or intervention taken under the current boundary model and current values (or explicit assumptions) of unresolved dimensions. Formally, let $\hat d_1$ , . . . , $\hat d_m$ be measured or assumed values for the active dimensions. Then $A=\pi(J,\Omega,\hat d_1,\ldots,\hat d_m)$, where π is a decision policy, human or machine. The role of ACTION is epistemic as well as practical: it commits the current representation to a real consequence. Without action or some equivalent test, the model cannot receive informative feedback.

\subsection{RESULT: confirming, refining, or expanding the model}

RESULT records evidence returned by the world after ACTION, with more structure than a success/failure label. A result can produce three distinct updates: 

\begin{enumerate}
\item \textbf{Boundary confirmation.} The result is consistent with the expected IS region.
\item \textbf{Boundary refinement.} The result falls on the BUT side, but the reversal can be explained using existing dimensions; a threshold or condition is revised.
\item \textbf{Dimension expansion.} The result cannot be adequately expressed using existing dimensions; a new candidate variable must be introduced and paired through a new IS/BUT analysis.
\end{enumerate}

The second and third cases should not be conflated. If $k/n$ = 0.08 performs poorly where the working boundary was 0.10, the scale dimension may remain adequate while its threshold moves. If the workload fails because mutation violates an ownership contract, no scale threshold can express the cause; the model requires a new dimension.

\section{The O-I-B-A-R Loop and Local Stability}

The complete loop can be summarized as $\mathrm{OPEN}\to(\mathrm{IS},\mathrm{BUT})\to D\to q_D\to\mathrm{ACTION}\to\mathrm{RESULT}\to\mathrm{OPEN}'$. The first OPEN is chosen. Subsequent OPEN states are often generated by the prior RESULT. A failed result may open a more specific question; a newly discovered variable may become the object of the next cycle.

\subsection{Structural fixed points are local, not completeness claims}

For a bounded decision problem and a specified stream of cases, call the current representation structurally stable when observed RESULT states no longer (i) expand the active dimension set or (ii) materially move the boundaries associated with existing dimensions. This is a property of the interaction between a representation and an observed case stream. It is not a claim that every decision-relevant dimension in the world has been discovered. Stability does not require all suspensions to disappear. Runtime variables may remain unresolved until a case occurs. A mature model may therefore contain stable dimensions with repeatedly instantiated suspensions. The relevant distinction is between ``we do not know the value yet'' and ``we do not know which variable matters.''

\subsection{A bounded exposure-rate certificate}

A run of cases with no new-dimension discovery should not by itself be interpreted as convergence. Rare dimensions produce long waiting times even when the representation remains incomplete. A simple stopping statement is available only under explicit sampling assumptions. Assume, for a monitoring window, that cases are sampled independently from a stationary exposure process and that each case reveals a previously unrepresented decision dimension with probability $p_{\mathrm{new}}$ . If T consecutive cases reveal no new dimension, then

\[
\Pr(0\ \text{discoveries}\mid p_{\mathrm{new}})=(1-p_{\mathrm{new}})^T.
\]

Consequently, a one-sided $(1-\delta)$ upper confidence bound is
\[
p_{\mathrm{new}}\le 1-\delta^{1/T}\le \frac{\ln(1/\delta)}{T}.
\]

This bound says only that, under the stated exposure process, the probability that the next sampled case exposes a new dimension is unlikely to exceed the bound. It does not certify completeness of the latent dimension set. The statement becomes invalid when the case distribution shifts, observations are strongly dependent, exposure is selectively filtered, or failure incidents are not reported. The last condition matters especially for organizational AI. A declining discovery rate may indicate a maturing representation, but it may also indicate declining candor. We call the latter silence-induced false stability: the model appears to have stopped generating new dimensions because the social process supplying informative BUT cases has deteriorated. Section 9 treats this as a deployment problem in its own right, not as a statistical footnote.

\section{Diagnostics: When the Scaffold Fails Informatively}

Inability to complete an \oibar{} field provides diagnostic information; the framework does not treat it as a formatting failure.

\begin{table}[htbp]
\centering
\footnotesize
\begin{tabularx}{\textwidth}{@{}YYY@{}}
\toprule
\textbf{Observed pattern} & \textbf{Likely diagnosis} & \textbf{Next move} \\
\midrule
IS empty, BUT empty & No operational knowledge or malformed OPEN & Find a concrete judgment and experienced cases \\
IS fluent, BUT empty & Domain too broad; imported slogan or definition & Narrow the domain until failures become recallable \\
IS empty, BUT rich & Experience without positive model & Work backward from failures to success conditions \\
BUT is only the opposite of IS & Zero new explanatory information & Ask for a concrete incident and mechanism \\
Multiple variables change between pair & Mixed dimensions & Split into separate boundary pairs \\
Suspension is ``it depends'' & Unknown is not yet typed & Ask what fact would decide the case \\
Dimension names become ``factor/attribute/feature'' & Abstraction lost value-bearing type & Narrow domain; return to the suspension question \\
Suspension cannot be pre-resolved & Runtime human judgment is required & Define the role by the variable being judged \\
\bottomrule
\end{tabularx}
\end{table}
The pattern ``fluent IS, empty BUT'' merits attention. Fluent rules are often rewarded as signs of clarity, yet a rule that produces no concrete failure cases may simply be a decontextualized norm. Narrowing the domain can turn generic prose into a falsifiable decision model.

\section{Worked Examples}

\subsection{Customer support: from empathy to need type}

\textbf{OPEN.} For renewal customers disputing a price or refund, emotional acknowledgment should precede solution discussion.

\textbf{IS.} When the customer is visibly upset and has not yet articulated the concrete remedy sought, acknowledgment often improves the interaction.

\textbf{BUT.} A customer repeatedly asks for a refund amount; the agent continues empathic statements; the customer becomes more frustrated and says that acknowledgment is not the requested outcome.

\textbf{Dimension discovery.} The action (empathy) is constant. The relevant changing variable is what kind of response the customer currently needs. The failure is attributed to misclassification of need type; empathy is the action held constant.

\textbf{Suspension.} Does this customer currently need to be heard, or do they need a concrete remedy or number?

\textbf{ACTION.} If the value is ``concrete remedy,'' move directly to amount, conditions, and timing.

\textbf{RESULT.} If this resolves the interaction, the boundary is supported. If the customer rejects the amount because similar customers received more, the existing dimension may be insufficient and a new candidate such as fairness perception may be required. This example illustrates why negative boundaries can improve human--AI handoffs. The machine does not need the vague rule ``escalate complex cases.'' It can escalate when the active dimension cannot be assigned a reliable value.

\subsection{Top-$K$ selection: dimension growth without a predefined taxonomy}

Suppose a team begins with a size-based rule for choosing a heap implementation.

\textbf{Boundary pair 1: scale.}

\begin{itemize}
\item IS: when $k/n$ is small, heap-based selection is attractive.
\item BUT: when k approaches a large fraction of n, the advantage disappears.
\item Suspension: how large is k relative to n?
\item Dimension: scale.
\end{itemize}

\textbf{Boundary pair 2: input form.} A new case reveals that the input arrives as a stream, whereas the initial case assumed a complete array. The relevant suspension becomes ``is the input batch or stream?'' and the dimension is input form.

\textbf{Boundary pair 3: ownership.} A further result shows that in-place heapification mutates data still used by the caller. The suspension becomes ``who owns this object and may it be mutated?'' and the dimension is ownership.

\textbf{Boundary pair 4: copying.} A memory regression raises ``does this implementation make a copy?'' leading to copying.

\textbf{Boundary pair 5: resource budget.} A production limit raises ``what peak memory is acceptable?'' leading to resource budget.

The resulting dimensions --- scale, input form, ownership, copying, and resource budget --- need not be listed before analysis. They emerge from separate failure boundaries. The framework therefore generates dimensions from cases and does not begin from a taxonomy.

\subsection{Research design: when the contribution appears in the ``limitations'' section}

Consider a research claim: ``reducing class size improves learning outcomes.'' A generic limitation such as ``the sample may not generalize to rural schools'' is largely a scope disclaimer. A more productive BUT is a concrete case in which class size fell but learning did not improve because teaching practice remained optimized for large classes, or because rapid hiring lowered teacher quality. The new terms --- teaching practice and teacher quality --- may point to dimensions absent from the original theoretical frame. In many research workflows, such variables are discovered late and relegated to a discussion section because the framework was fixed earlier. \oibar{} uses such negative evidence for construct formation during the research process, so its role is no longer confined to post-hoc qualification.

\section{Intellectual Lineage and Relationship to Prior Work}

Several parts of \oibar{} have clear ancestors. Locating the framework in that lineage supports a more defensible claim: its contribution lies in a particular representation and coupling of mechanisms. Contrastive learning and knowledge elicitation themselves are established ideas.

\subsection{Situated action, organizational routines, and noncanonical practice}

The cs.CY motivation begins with work showing that formal organizational representations underdescribe situated action. Suchman argued that plans function as resources for action but do not determine the concrete course of situated activity \cite{r5}. Brown and Duguid showed that actual work practices can differ fundamentally from manuals and formal job descriptions, with learning and innovation occurring in noncanonical practice \cite{r6}. Orr documented how service technicians use context-rich stories as a socially distributed resource for diagnosis beyond formal documentation \cite{r7}. Feldman and Pentland's ostensive/performative distinction similarly warns against equating a routine's abstract representation with its enacted performances \cite{r8}.

\oibar{} operationalizes one narrow implication for AI rule authoring: if breakdown cases carry knowledge that is absent from canonical rules, those cases should not remain merely anecdotal. They can be represented as negative boundaries and used to expose variables needed for future discrimination.

\subsection{Tacit knowledge, externalization, and AI-enabled knowledge transfer}

Polanyi's account of tacit knowing established the familiar asymmetry between what people can do and what they can readily state \cite{r15}. Nonaka's theory of organizational knowledge creation made externalization one mode in a dynamic relationship between tacit and explicit knowledge \cite{r16}. Recent work shows renewed computational interest in this problem. Cho et al. report that LLMs can extract useful tacit knowledge from expert behavior and improve downstream transfer \cite{r17}; PRAXIS similarly treats undocumented business rules, interface contracts, and operational conventions as a limiting factor for domain code-generation agents \cite{r18}.

These works strengthen the motivation for externalization but leave open the target representation. For organizational decisions, \oibar{} combines prose with bounded judgments, value-bearing dimensions, unresolved value queries, and feedback rules.

\subsection{Kelly's repertory grid: contrast without a predefined taxonomy}

Kelly's personal construct theory and Role Construct Repertory Test provide an early and important precedent for eliciting dimensions through contrast without a fixed questionnaire taxonomy \cite{r9}. In the classic triadic elicitation, a respondent considers three elements, identifies how two are alike and different from the third, and thereby produces a bipolar construct. Repertory-grid methods later became a significant knowledge-acquisition technique for expert systems: interactive systems elicited, analyzed, and refined expert construct systems, and tools such as the Expertise Transfer System used personal construct psychology for knowledge-base construction \cite{r13,r14}.

\oibar{} makes no broad novelty claim for ``taxonomy-free dimension elicitation.'' The methods differ in the source and semantics of their contrasts. Kellyian elicitation contrasts selected elements to reveal a person's construing. \oibar{} contrasts an applicability case with a consequential failure of a judgment. Its contrast is outcome-asymmetric, while Kellyian contrast is organized around similarity. The output is also coupled to a suspension: a query for the value that must be resolved before acting in the next instance. Future RESULTs can expand the dimension set beyond populating or reorganizing a current grid.

\subsection{Winston's near misses: the historical ancestor of the one-difference heuristic}

Winston's 1970 dissertation and its 1975 chapter version are the closest classical AI ancestors of the boundary-pair idea \cite{r10,r11}. Winston defined a near miss as a training sample that is very similar to the target concept but differs from it in only a small number of significant points. Carefully constructed positive and negative examples allow the learner to identify which relations are essential to a concept.

The \oibar{} target |∆(IS, BUT | Ω)| = 1 operationalizes the near-miss intuition for messy organizational elicitation; it is not a new principle of learning from one distinguishing feature. Winston assumes a teacher who deliberately constructs a useful training sequence in a small, represented blocks world. \oibar{} begins from recalled incidents whose descriptions may mix many variables and asks the analyst to decompose them until a single decision-relevant change can be examined.

The output also differs. Winston's program learns structural constraints for concept membership. \oibar{} seeks a decision dimension whose value may vary across future cases, then explicitly represents the intermediate state in which the dimension is known but its current value is not. That suspension state, and its use for human--AI handoff, has no direct analogue in Winston's concept learner. Finally, Winston's concept vocabulary is effectively closed by the scene representation and teaching environment; \oibar{} allows RESULT to expose variables absent from the current dimension set.

\subsection{Mitchell's version spaces: positive and negative examples within a fixed hypothesis language}

Mitchell's candidate-elimination algorithm maintains the set of hypotheses consistent with positive and negative examples and refines the specific and general boundaries of a version space \cite{r12}. This work reinforces a general point shared with \oibar{}: negative examples supply information that can specialize a model; they are more than error labels.

The difference is representational. Version-space learning presupposes an instance representation and a hypothesis language in which the target concept is to be expressed. \oibar{} is aimed at situations where the organization may not yet know the relevant decision variable or even the vocabulary in which it should be stated. Its boundary pair therefore helps construct the feature/dimension representation that a conventional learner would normally receive as input.

\subsection{Cognitive task analysis and the Critical Decision Method}

CTA and the Critical Decision Method use retrospective probing of nonroutine incidents to elicit expert cues, discriminations, situation assessments, and decision requirements \cite{r19,r20}. In practical knowledge elicitation, these methods are arguably closer to \oibar{} than abstract concept learning because they begin from consequential events in expert work. The methods differ mainly in their target representations; both use incidents. CTA may yield timelines, cue inventories, knowledge audits, and decision requirements. \oibar{} imposes a particular transformation: pair a success region with a concrete failure, decompose the contrast

\begin{table}[!htbp]
\centering
\scriptsize
\setlength{\tabcolsep}{3pt}
\begin{tabularx}{\textwidth}{@{}p{0.11\textwidth}YYYY@{}}
\toprule
\textbf{Tradition} & \textbf{Source of contrast / evidence} & \textbf{Representation assumptions} & \textbf{Typical output} & \textbf{What OIBAR adds or changes} \\
\midrule
Kelly repertory grid & selected triads; two alike, one different & constructs elicited from respondent; no fixed questionnaire taxonomy & bipolar personal constructs and grid & outcome-asymmetric success/failure incidents; suspension and responsibility; open RESULT loop \\
Winston near miss & teacher-curated positive/negative examples & represented blocks world and comparison vocabulary & concept membership constraints & recalled incidents; decomposition of messy failures; decision dimensions; typed unknown values \\
Mitchell version spaces & labeled positive/negative instances & predefined instance and hypothesis language & boundaries of consistent hypotheses & uses contrast partly to construct missing decision representation \\
CTA / CDM & retrospective critical incidents & expert interviews and probe structure & cues, timelines, decision requirements & fixed transformation into boundary pair, dimension, suspension, update rule \\
CEGAR & formal counterexample traces & formal transition system / abstraction & refined formal abstraction & organizational incidents, open vocabulary, role/handoff semantics \\
Situated-practice research & ethnography of actual work & no computational representation required & account of canonical/noncanonical practice & operationalizes breakdowns as reusable decision-boundary structures \\
\bottomrule
\end{tabularx}
\caption{OIBAR in relation to major intellectual ancestors and adjacent methods.}
\label{tab:lineage}
\end{table}
toward a value-bearing variable, convert the current unknown value into a suspension, and let later results either modify the boundary or grow the variable set.

\subsection{\texorpdfstring{Boundary conditions, double-loop learning, and\\counterexample-guided refinement}{Boundary conditions, double-loop learning, and counterexample-guided refinement}}

Research on theoretical boundary conditions emphasizes the ``who, where, and when'' limits under which theoretical relationships hold \cite{r21}. \oibar{} shares the concern with scope and extends the role of boundary contrast: it generates candidate variables in addition to supplying metadata around an existing theory.

Double-loop learning distinguishes changing actions from questioning the governing assumptions behind them \cite{r22}. \oibar{} offers a local representational mechanism for a related move: a RESULT that existing dimensions cannot explain triggers examination of the dimension set itself.

At a more distant computational level, CEGAR uses counterexamples to iteratively refine an abstraction in formal verification \cite{r23}. The analogy is useful but limited. CEGAR operates over formally specified systems and counterexample traces; \oibar{} works over human incident descriptions, contested judgments, and open organizational vocabularies. Its refinement target may add a newly named decision dimension and a new human responsibility alongside a more precise formal abstraction.

This genealogy narrows the novelty claim. Contrastive elicitation, near-miss learning, and positive/negative boundary refinement are established ideas. The proposed contribution is their synthesis around an organizationally specific object: an open set of decision dimensions whose unresolved values can define human--AI coordination and whose composition can change through feedback.

\section{Sociotechnical Implications for Human--AI Organizations}

\subsection{AI rules need explicit negative boundaries}

A common AI instruction is an abstract norm: ``be professional,'' ``show empathy,'' ``follow the refund process,'' or ``escalate special cases.'' Such instructions ask a model to infer boundaries that the organization has not represented. \oibar{} suggests that a machine-usable organizational rule should ideally contain at least three pieces of information:

\begin{enumerate}
\item when the action is justified (IS),
\item a verified or testable failure boundary (BUT), and
\item what fact must be established before selecting a side (suspension).
\end{enumerate}

Complete enumeration of every exception is unnecessary. The rule needs enough negative evidence to identify the dimension on which a future case turns.

\subsection{From generic ``human in the loop'' to variable-specific responsibility}

The suspension representation reframes human oversight. If the missing value can be retrieved or measured, the system should obtain it. If the value can only be produced through situated interpretation, then the organization has identified a concrete reason for human involvement. This yields a more specific handoff contract:

\[
\text{handoff}\iff\text{a decision-critical value cannot be resolved reliably}.
\]

The variable being judged describes the human role alongside the job title. A customer-service agent may be required because the current need type cannot be inferred reliably; a senior engineer may be required because the current failure cannot be expressed by known dimensions and must be classified as threshold error versus model expansion.

This view complements organizational research on human--AI augmentation \cite{r2,r3}: the boundary between human and machine is not necessarily a fixed task partition. It can be a dynamically located information-acquisition boundary.

\subsection{AI assistance is safest where its outputs are independently checkable}

The framework suggests an epistemic division of labor. AI-generated IS statements are risky when IS represents the user's substantive claim; a fluent model can invent a plausible theory and make it difficult to distinguish the organization's knowledge from generated prose. By contrast, AI can be more active where output can be checked against records: proposing candidate counterexamples, asking for missing incident details, detecting multiple variables mixed in one BUT, rewriting ``it depends'' as an answerable suspension, and proposing candidate dimension names. A useful heuristic is:

\begin{quote}
\emph{AI proposes counterexamples and questions; humans own claims, incident validity, and result interpretation.}
\end{quote}

Recent work showing that LLMs can extract tacit expertise from behavioral traces \cite{r17} makes this division increasingly practical, while the \oibar{} structure provides an auditable target for what is being extracted.

\section{The Politics of Externalizing Failure Knowledge}

The preceding sections describe why a structured history of BUT cases can support organizational learning. In a sociotechnical system, however, making failure knowledge inspectable is not an unqualified benefit. It changes the social object being recorded.

\subsection{From war story to attributable record}

Orr's copier technicians exchanged diagnostic stories in situated peer practice outside the official repair documentation \cite{r7}. Such stories could preserve the contextual detail that canonical manuals omitted, while remaining embedded in a local social setting. An OIBAR deployment may transform the same kind of material into a durable record: a dated failure, a named decision dimension, an action, and a result. That transformation can improve reuse, but it can also alter incentives to disclose.

Organizational-silence research argues that employees may withhold information about problems when speaking up is perceived as risky or futile \cite{r24}. Psychological safety similarly concerns whether interpersonal risk taking is experienced as safe enough to support learning behavior \cite{r25}. Tucker and Edmondson show how front-line problem solving can restore local work while leaving underlying causes unaddressed, thereby inhibiting organizational learning from failure \cite{r26}. These literatures imply that a failure-elicitation method cannot assume a neutral supply of candid incidents.

This creates a tension:

\begin{quote}
\emph{OIBAR needs truthful failure histories to make organizational knowledge legible to AI, but making those histories durable, attributable, and machine-readable may reduce people's willingness to produce them.}
\end{quote}

This tension directly affects the framework. It changes the observed BUT distribution and can distort which dimensions appear to matter.

\subsection{Applying OIBAR to its own externalization claim}

The method can expose this tension using its own structure:

\begin{table}[htbp]
\centering
\small
\begin{tabularx}{\textwidth}{@{}p{0.16\textwidth}Y@{}}
\toprule
\textbf{Element} & \textbf{Self-application} \\
\midrule
OPEN & Externalizing failure knowledge improves organizational learning and AI rule quality. \\
IS & When contributors can report failures without disproportionate personal cost, structured records make decision boundaries reusable. \\
BUT & When failure records are durable and personally attributable, contributors may withhold, sanitize, or depersonalize precisely the incidents from which the organization most needs to learn. \\
Suspension & Can this incident be recorded with enough contextual detail to preserve the decision variable without imposing unacceptable attribution exposure on a contributor? \\
Candidate dimension & Personal attribution exposure: to what extent can the failure record be traced to a specific individual and linked to consequential evaluation? \\
\bottomrule
\end{tabularx}
\end{table}
The framework does not imply universal anonymity or the removal of accountability. Accountability, privacy, candor, due process, and learning can conflict. The record architecture itself becomes part of the decision problem.

\subsection{Design questions for organizational deployment}

A serious implementation has to specify the governance of OIBAR fields along with their content. Open design questions include:

\begin{itemize}
\item whether raw incident narratives are separated from reusable boundary representations;
\item whether identity is necessary for learning, accountability, both, or neither in a given use case;
\item who may read raw BUT histories versus validated dimension-level abstractions;
\item whether learning records are institutionally separated from performance or disciplinary records;
\item how corrections, contestation, and alternative interpretations of a failure are preserved;
\item how organizations detect silence-induced false stability and avoid interpreting an empty incident stream as evidence that the model has converged.
\end{itemize}

These are not implementation details that can be solved by the representation alone. They are part of the cs.CY problem: the same technical act of externalization can redistribute visibility, accountability, and risk inside an organization.

\section{Representational Sanity Check: Demonstrating State Aliasing}

No synthetic accuracy comparison is needed to establish the most basic representational point. Consider an IS-only representation $\phi_{\mathrm{IS}}(x)$ that records whether the documented positive condition holds. Let $x_B$ be a BUT case that requires an alternative action and $x_U$ a suspension case that requires querying or escalation. If both cases are simply represented as ``not IS,'' then

\[
\phi_{\mathrm{IS}}(x_B)=\phi_{\mathrm{IS}}(x_U).
\]

Any deterministic policy that conditions only on $\phi_{\mathrm{IS}}$ must therefore choose the same behavior for both cases, even though the required behaviors differ. The representation aliases two organizationally distinct states.

This modest proposition establishes a representational limitation. If real organizational practice distinguishes negative-boundary cases from unresolved-value cases, a one-sided rule representation that collapses them cannot preserve that distinction by construction. The proposition makes no performance claim for real AI systems. The empirical question is whether real incidents yield stable and decision-relevant BUT states, dimensions, and suspensions, and whether representing them improves human--AI work relative to less structured alternatives.

The source package contains a minimal executable check that encodes this state-aliasing proposition. It is included only to confirm that the software representation matches the stated construction; no synthetic performance percentages are reported or treated as evidence.

\section{Testable Hypotheses and Evaluation Agenda}

The framework should be treated as falsifiable. Its value depends on whether the proposed representational distinctions improve real organizational work.

H1: Boundary robustness. Given the same source expertise, rules authored with explicit IS, concrete BUT, and suspension queries will produce fewer inappropriate actions in boundary cases than one-sided positive rules.

H2: Dimension discovery. On domains containing unanticipated failure causes, boundary-first elicitation will identify more decision-relevant dimensions absent from the initial rule set than a fixed-category checklist.

H3: Handoff specificity. Suspension-based human handoffs will reduce both missed escalations and unnecessary defensive escalations relative to vague criteria such as ``complex cases.''

H4: Transfer. Novices receiving dimension-level failure attribution (``need type was misclassified'') will transfer learning to novel cases better than novices receiving action-level admonitions (``do not over-empathize'').

H5: Organizational learning. Incident reviews that distinguish threshold revision from new-dimension discovery will produce more reusable model updates and fewer action-specific guardrails than conventional postmortems.

\textbf{Stability calibration.} The stopping rule in Section 4 is not an OIBAR-specific performance hypothesis. It is a statistical monitoring statement under explicit exposure assumptions. An empirical study can nevertheless test whether its nominal upper bounds are calibrated under a pre-registered, approximately stationary case-sampling regime, and separately measure whether reporting incentives alter the observed incident stream.

\subsection{Minimal empirical design}

A first study need not be large. Collect a small corpus of real organizational rules and verified incidents from the people who use them. For each rule, construct:

\begin{itemize}
\item Baseline: the original one-sided rule;
\item OIBAR: the same rule plus verified BUT cases, dimension attribution, and suspension query;
\item optionally, taxonomy baseline: a version augmented using a predefined checklist of common dimensions.
\end{itemize}

Human participants or AI systems can then be tested on ordinary cases, held-out boundary cases, and cases where the critical value is deliberately withheld. Outcomes include inappropriate action rate, missed-question rate, unnecessary escalation, transfer to novel cases, analyst agreement on extracted dimensions, and the frequency with which held-out incidents introduce dimensions absent from the initial representation.

A second study can focus on knowledge elicitation itself: compare Kelly-style triadic elicitation, CTA/CDM interviewing, a fixed taxonomy, and OIBAR boundary-first elicitation on time cost, number of nonredundant dimensions, stability across analysts, and usefulness for downstream decisions. This comparison is necessary because ``eliciting dimensions without a taxonomy'' alone does not establish OIBAR's novelty.

The framework would be weakened if the added structure merely increases verbosity without improving decisions, if independent analysts cannot agree on stable dimensions, if suspensions do not predict useful questions or handoffs, or if fixed taxonomies perform equally well on novel failure causes.

\section{Limitations}

First, the ``one changing variable'' criterion is an operational decomposition target, not a descriptive claim that real organizational failures have a single cause. Winston's original near-miss notion itself allows a small number of significant differences \cite{r11}. OIBAR uses |∆| = 1 to force mixed incidents to be decomposed into analyzable boundary pairs; whether a proposed decomposition is causally or decision-relevantly adequate remains a human judgment.

Second, Kelly, Winston, Mitchell, CTA/CDM, and CEGAR provide precedents for contrastive elicitation, taxonomy-free construct elicitation, positive/negative example learning, and refinement guided by counterexamples. The empirical burden falls on the specific integration proposed here: open-world dimension generation from organizational incidents, suspension as a typed unknown and handoff primitive, domain-relative naming, and expansion of the represented decision space through RESULT.

Third, the presence of new vocabulary in a BUT is only a heuristic for generativity. A sentence can introduce irrelevant words, and a genuine new dimension may sometimes be expressed using existing vocabulary. The substantive criterion is explanatory and decision relevance to the boundary transition.

Fourth, domain-relative naming can be unstable when the domain is underspecified or when multiple professional vocabularies overlap. The fixed-point notation is an operational description of a naming procedure, not a proof of unique semantic convergence.

Fifth, a RESULT does not automatically reveal whether a threshold should move or a new dimension should be introduced. That distinction is itself a high-value expert judgment and may remain contested.

Sixth, apparent representational stability can be a sampling artifact. Rare dimensions, distribution shift, selective observation, or suppressed failure reporting can all produce long periods with no new dimensions. The exposure-rate bound in Section 4 is conditional and cannot be interpreted as a completeness guarantee.

Seventh, externalization changes incentives. Durable failure records may increase learning while also increasing attribution exposure, interpersonal risk, or incentives to sanitize incidents. Organizational silence and psychological safety directly affect the data-generating process from which BUT cases are elicited.

Finally, not every form of tacit expertise can or should be fully externalized. Some expertise is embodied, socially distributed, normatively contested, private, or costly to articulate. The intended claim is narrower: when organizational expertise can be expressed through decision-relevant success and failure boundaries, \oibar{} provides a structured route from incidents to explicit dimensions and runtime questions, subject to governance arrangements that preserve both accountability and candid learning.

\section{Conclusion}

The organizational problem addressed in this paper extends beyond the intelligence of AI systems or the volume of documents available to them. Formal organizational knowledge is itself a partial representation of work. Manuals, policies, prompts, and procedures tend to preserve the ostensive positive rule, while experienced practice also relies on negative boundaries, live contextual values, responsibility for judgment, and accumulated failure history.

\oibar{} proposes one way to make that missing structure explicit. It asks six connected questions: What is the rule? Where does it work? Where has it concretely failed? What changed between those states? What value must be known before acting now, and who can obtain it? What does the next result force us to revise? The resulting representation is an evolving set of value-bearing decision dimensions and their boundaries, not an expanded rule alone.

The framework has clear ancestors. Kelly showed that constructs can be elicited from contrast without imposing a fixed questionnaire taxonomy. Winston showed the learning power of carefully chosen near misses. Mitchell formalized positive/negative boundary refinement within version spaces. CTA/CDM showed how critical incidents reveal expert cues and decision requirements. Situated-action and organizational-routine research showed why formal descriptions of work cannot be equated with actual practice. \oibar{} should be judged as a proposed synthesis for a new deployment context that builds on this lineage.

Its most consequential sociotechnical claim may be the suspension. When the dimension is known but its current value is not, the system knows what it does not know. If the value can be retrieved, the machine can retrieve it; if it can only be produced through situated judgment, the unresolved variable defines why and where a human is needed. This turns ``human in the loop'' from a generic safety slogan into a decision-specific responsibility. The proposal is:

\begin{quote}
\emph{Use organizational failure to discover the decision axis that a rule failed to represent, then record what must be known before that axis can guide action again.}
\end{quote}

Whether this representation improves real human--AI systems is an empirical question. The framework's value will be determined by organizations' ability to transfer expert judgment more faithfully, make human handoffs more specific, and learn from deployment failures without accumulating another layer of brittle rules. Formal elegance alone cannot establish that value.

\appendix
\section{A Minimal O-I-B-A-R Elicitation Template}

\begin{enumerate}
\item OPEN. State one judgment that can fail, and the concrete domain in which it will be tested.
\item IS. Under what condition is the judgment currently expected to hold?
\item BUT. Recall a concrete case in which applying the judgment failed. Do not write the logical opposite of IS.
\item Compare. What decision-relevant variable changed? If none changed, rewrite the BUT. If several changed, split the incident into multiple candidate boundary pairs.
\item Attribute. Rewrite the failure as a judgment error (``we misjudged d'') and treat the action error as a secondary description.
\item Suspend. What fact must be established to determine the value of d in the present case?
\item Name. Remove instance details and recursively ask within the domain what the suspension question is about until a value-bearing term stabilizes.
\item ACTION. Measure, retrieve, ask, or explicitly assume the current value and act.
\item RESULT. Did the outcome confirm the boundary, move a threshold, or reveal a variable the current model cannot express?
\end{enumerate}

\section{Executable Representational Check}

The source package includes representational\_sanity\_check.py. The script contains no benchmark or synthetic accuracy score. It encodes three states---IS, BUT, and suspension---and asserts two properties: an IS-only encoding aliases the latter two states, while an OIBAR-style encoding can preserve their distinction. The script is an executable statement of the proposition in the main text, not empirical evidence.


\end{document}